\documentclass{article}\usepackage{jkas2}

\runningauthor{Dinshaw Balsara}
\runningtitle{ ADAPTIVE MESH REFINEMENT}

\begin{document}

\title{ADAPTIVE MESH REFINEMENT IN COMPUTATIONAL ASTROPHYSICS -- METHODS
AND APPLICATIONS}

\author{DINSHAW BALSARA}

\address{ Department of Physics, University of Notre Dame, Notre Dame, 
IN 46556, USA}

\address{\normalsize{\it (Received  Nov 23, 2001; Accepted Nov. 26, 2001)}}

\abstract{
The advent of robust, reliable and accurate higher order Godunov 
schemes for many of
the systems of equations of interest in computational astrophysics
has made it important to understand how to solve them in multi-scale
fashion. This is so because the physics associated with 
astrophysical phenomena evolves in multi-scale fashion and we 
wish to arrive at a multi-scale simulational capability to represent
the physics. Because astrophysical systems have magnetic fields,
multi-scale magnetohydrodynamics (MHD) is of especial interest. In this
paper we first discuss general issues in adaptive mesh refinement (AMR).
We then focus on the important issues in carrying out divergence-free
AMR-MHD and catalogue the progress we have made in that area. We show
that AMR methods lend themselves to easy parallelization. We then discuss
applications of the RIEMANN framework for AMR-MHD to problems 
in computational astophysics.} 


\maketitle

\section{INTRODUCTION}

Observations of various astrophysical systems such as proto-stars,
novae, supernovae, the interstellar medium, galaxies and cosmology
show that physical processes in these systems evolve in multi-scale
fashion. To take but a single example from proto-star formation,
once a Class 0 core forms out of the turbulent, magnetized, molecular
gas in a molecular cloud, gravity causes the system to be largely decoupled
from the rest of the turbulent flow. And yet, the decoupling is not
complete. Large-scale magnetic fields cause the proto-star's angular
momentum to be coupled to that of the external medium. Likewise,
different parts of the dusty protostellar envelope are radiatively
coupled to each other. The radiative coupling also sets the temperature
of the gas, thereby determining its coupling with the magnetic field. 
As a result, we see the need for: (1) detailed representation of
the micro-physical processes, (2) accurate representation of the system
of equations that are of interest in computational astrophysics and
(3) an ability to do (1) and (2) in a multi-scale fashion via adaptive
mesh refinement (AMR). AMR is an elegant technique for concentrating mesh
resolution in regions where an accurate answer is desired. Almost all
astrophysical processes have magnetic fields and so an ability to
represent MHD in multi-scale fashion is a {\it sine qua non} for 
computational astrophysics. The present paper focusses on recent
advances by the author that have made it possible to carry out
AMR-MHD calculations.

Ever since the first paper on AMR methods for fluid dynamics, see 
Berger and Colella (1989), it has been recognized that that higher
order Godunov schemes play a central role in practical AMR simulations.
The reasons are not difficult to see and the talks by Ryu (this conf.)
on MHD, Ibanez and Koide (again this conf.) on relativistic flow
bear testimony to the usefulness of these techniques in 
computational astrophysics. Such schemes were first formulated 
for hydrodynamics by vanLeer (1979), where their usefulness 
has been very well-documented by Woodward and Colella (1984).
In recent years, they have been formulated for several other
systems of interest in computational astrophysics wich include
relativistic hydrodynamics, see Balsara (1994); radiation
hydrodynamics, see Balsara (1999a,b,c); radiation MHD, see
Balsara (1999d,e); relativistic MHD, see Balsara (2001a) 
and multidimensional radiative transfer, see Balsara (2001b).
Perhaps the most vigorous evolution has taken place in MHD
where Roe and Balsara (1996) designed the first complete MHD
eigenvectors that were free of singularities; Brio and Wu
(1988), Zachary, Malagoli and Colella (1994), Powell (1994), Dai and
Woodward (1994), Ryu and Jones (1995) and Balsara (1998a) designed
Riemann solvers for MHD; Dai and Woodward (1995), Ryu et al (1998) 
and Balsara (1998b) catalogued different forms of TVD schemes for
MHD and Balsara and Spicer (1999), Dai and Woodward (1998), 
Ryu et al (1998), Londrillo and Del Zanna (2000) and Toth (2000)
formulated divergence-free higher order Godunov schemes for MHD.
The latter divergence-free formulations realize that the divergence
of the magnetic field should remain exactly zero. 
Brackbill and Barnes (1980) and Brackbill (1985)
have shown that violating the
constraint leads to unphysical plasma transport orthogonal to the magnetic
field as well as a loss of momentum and energy conservation.
Powell et al (1999) did indeed formulate an AMR
scheme for MHD that was not divergence-free only to
find that the maximal errors ocurred on the finest meshes due to
unphysical build-up of divergence. The finest meshes are the
very meshes where one would have wanted the error to be minimal!
Moreover, in accreting astrophysical flows, the divergence would flow with
the fluid and build up in the very regions where one wants the
most accurate answer!
This prompted the present author to realize that only a
scheme that was divergence-free on the entire AMR hierarchy would
be adequate for astrophysical applications. These advances have been
catalogued in detail in Balsara (2001c) and implemented in
the RIEMANN framework for computational astrophysics.
While Balsara (2001c) provides more mathematical details, the
present paper focusses more on the intuitive ways of thinking
about the subject. As a result, the two papers complement
each other.

In Section II we discuss algorithmic issues in AMR hydrodynamics.
In Section III we discuss divergence-free AMR-MHD. In Section IV
we show how AMR-MHD techniques have been parallelized. In Section V
we discuss tests and applications.

\section{AMR HYDRODYNAMICS}

\begin{figure*}[t]
\centerline{\epsfysize=8cm\epsfbox{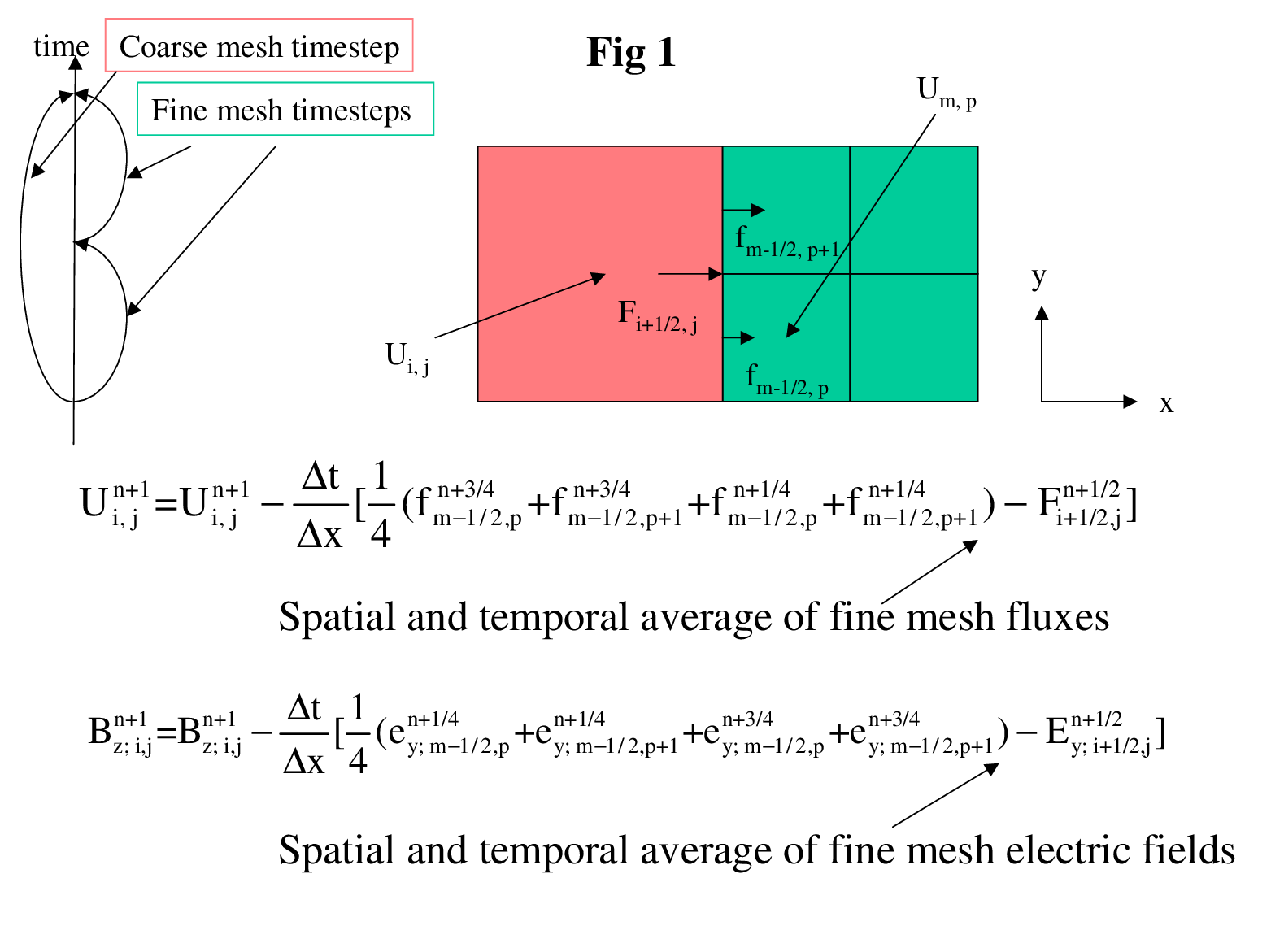}}
\vskip -1.0cm
\end{figure*}

Fig. 1 provides a schematic of a two-dimensional AMR calculation where we show
a large mesh zone and an adaptively refined set of four finer mesh
zones adjacent to it. The large mesh zone should be viewed as being
part of a coarse mesh and the four finer mesh zones should be viewed
as being part of a fine mesh in the AMR hierarchy that abuts the
above-mentioned coarse mesh. A consideration of that figure allows us to
motivate the four most important algorithmic issues in 
AMR-hydrodynamics that 
were formulated by Berger and Colella (1989) (hereafter BC). They are:

{\bf (II.a) Time step sub-cycling on refined meshes:} A look at Fig. 1
shows that the finer mesh has zones that are half as small as the
coarse mesh. As a result, the Courant condition for time step control
decrees that the finer mesh can only evolve with time steps that are
half as small as the coarse mesh time steps. This is illustrated in
the schematic time-axis in Fig. 1. As a result of this limitation,
BC realized that the fine mesh should move with a time step that is
an integral fraction of the coarse mesh time step. The precise fraction
is determined by the mesh refinement ratio. As a result, the fine mesh
will undergo more than one time steps before it reaches time-synchonization
with the coarse mesh. 

{\bf (II.b) Prolongation of coarse mesh solution to fine mesh boundaries/or new
fine mesh interiors:}
As the fine mesh undergoes fractional time steps, it will receive 
temporally interpolated boundary information from the abutting
coarse mesh. Even when a new fine mesh is built in a portion of
the coarse mesh that was not covered by a fine mesh, one has to
find a {\it conservative} strategy for transferring the solution from
the coarse mesh to the fine mesh. This step is known as prolongation. 
To prolong the solution from the coarse mesh to the fine mesh
one has to use the conservative interpolation of the 
underlying higher order Godunov scheme.

{\bf (II.c) Flux correction at the fine-coarse interface:} When the
fine mesh and coarse mesh solutions are synchronized, we want to
ensure conservation of mass, momentum and energy. As pointed out
by BC, a failure to ensure full conservation will result in spurious
solutions on the entire AMR hierarchy. This is especially true for
most astrophysical systems which give rise to very deep AMR hierarchies,
thereby exacerbating the problems associated with a loss
of conservation. This ability to ensure
conservation is only available when using a higher order Godunov
scheme as the underlying solution technique. For example, the staggered
mesh formulation of the ZEUS scheme of Stone and Norman (1991) cannot
ensure conservation of all variables on AMR hierarchies, thereby 
limiting its utility for AMR simulations. The flux conservation can
be imposed at the times when the fine and coarse meshes are synchronized.
This happens at times $t^n$ and $t^{n+1}$ where
the time step is given by $\Delta t = t^{n+1} - t^n$ .
The flux conservation is carried out by replacing 
the coarse mesh flux $F_{i+1/2,j}^{n+1/2}$ by a 
spatial and temporal average
of the fine mesh fluxes, i.e. $f_{m-1/2,p}^{n+1/4}$ , $f_{m-1/2,p+1}^{n+1/4}$ ,
$f_{m-1/2,p}^{n+3/4}$ and $f_{m-1/2,p+1}^{n+3/4}$ , at the interface 
between the fine and coarse mesh as shown in Fig. 1. As a result,
a layer of coarse mesh zones that abut the fine mesh zones will have
to undergo this flux correction. The variable $U_{i,j}^{n+1}$ will,
therefore, undergo flux-correction, as shown in Fig. 1, whenever
the fine and coarse meshes synchronize in time.

{\bf (II.d) Restriction of fine mesh solution to coarse mesh:} When
the fine and coarse meshes are time synchronized, the fine mesh
holds the more accurate solution. As a result, we replace the coarse
mesh solution by the spatially averaged fine mesh solution. 
This step is known as restriction.

This completes the process of demonstrating the four essential steps
in BC's AMR strategy for hydrodynamics. 
These four steps yield a scheme that is fully {\it conservative} of
mass, momentum and energy.
Notice that the coarse meshes deliver their
data to the fine mesh in step (b) above. Likewise, the fine meshes
deliver data that is generated on them to modify the coarse mesh
data in steps (c) and (d). As a result, the solution that evolves
on an AMR hierarchy is intimately connected across the levels in
the AMR hierarchy! This ensures that all the levels in the AMR hierarchy
are causally connected. It also ensures that a spurious solution generated
on one level in an AMR hierarchy (for example, the finest level in
the scheme of Powell et al (1999)) will eventually propagate and corrupt
the solution at all levels in the AMR hierarchy. This brings 
out the very interesting fact
that AMR techniques are physically consistent at a very deep level.
At the same time, it shows that violating that consistency can result
in a solution that is flawed at all levels (all puns intended!).

\section{DIVERGENCE-FREE AMR-MHD}

\begin{figure*}[t]
\centerline{\epsfysize=8cm\epsfbox{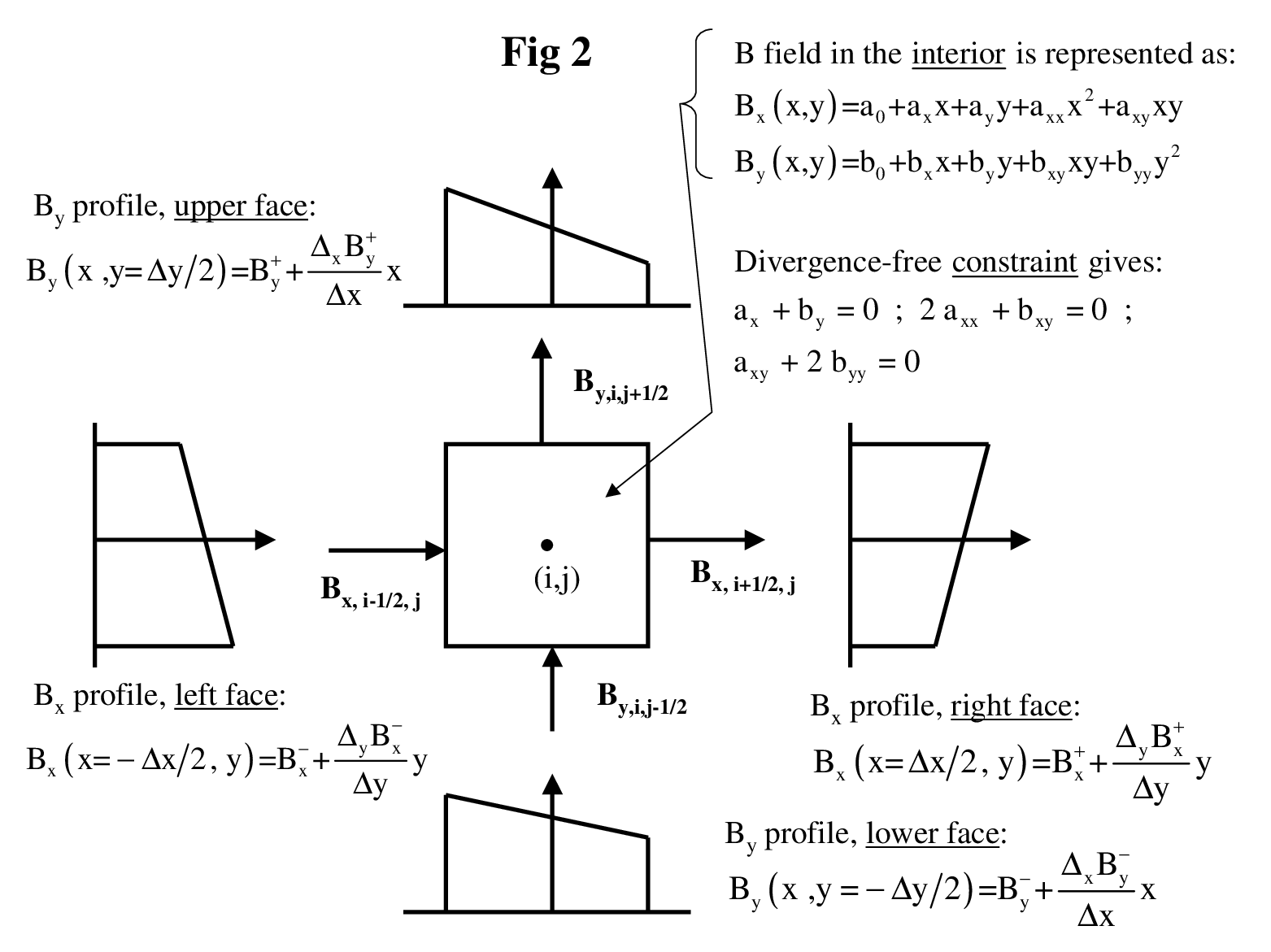}}
\vskip -0.5cm
\end{figure*}

The previous section suggests that it would be desirable to have
a similar strategy for AMR-MHD. However, a little thought reveals
a major obstacle. We saw in section II.b that it was essential
to use the reconstruction strategy for the underlying higher order
Godunov scheme in order to arrive at a conservative strategy
for AMR-hydrodynamics. However, up until 
recently, an analogous divergence-free
reconstruction strategy for (divergence-free) magnetic fields did not
exist. As a result, it was not possible to formulate divergence-free 
AMR-MHD because it was not possible to transfer the magnetic fields
in divergence-free fashion from a coarse mesh to a fine mesh.
In retrospect this lack of
a divergence-free reconstruction strategy might seem a little surprising since
the formulation of a reconstruction strategy is central to the 
process of designing a higher order Godunov scheme. 
Balsara and Spicer (1999) had, nevertheless, been able to design
a second order accurate Godunov scheme for divergence-free MHD
without having to address the reconstruction question. The reason
one is able to bypass the issue of divergence-free
reconstruction in the formulation of second order Godunov schemes
stems essentially from a mathematical anomaly. It turns out that
for second order schemes (and only for second order schemes) it is
possible to get by without having to address the issue of divergence-free
reconstruction. However, in order to formulate a divergence-free AMR-MHD
one has to face up to the the task of formulating a divergence-free
reconstruction strategy. We do that next for two-dimensions in this 
paper. Mathematical details associated with the two-dimensional
case as well as the analogous three-dimensional problem are catalogued
in Balsara (2001c).

Consider Fig. 2 which shows the four magnetic field components on the four
faces of a square. This is the lay-out of magnetic field variables
from Balsara and Spicer (1999) and yields
a scheme for divergence-free magnetic
field evolution. To arrive at a second order accurate formulation we
endow the field components with piecewise linear variation in the
transverse direction as shown in Fig. 2. 
One may well wonder why this should necessarily
yield a second order accurate formulation. It can be shown, see Balsara
(2001c), that for variations that are restricted to lie in one
dimension, this yields a second order accurate TVD scheme. 
Multidimensional TVD schemes can be build out of such one dimensional
building blocks.
As a result, we know that the strategy of endowing the field components
with piecewise linear variations in the transverse direction will
yield a second order accurate representation of the field in the
interior of the square. The most general polynomial representation of
the vector field in the interior of the square that matches the 
linear profiles at the boundaries is also given in Fig. 2. 
In order for the polynomials to be divergence-free the 
polynomial coefficients should satisfy the divergence-free
constraints, also given in Fig. 2. These constraints can be
straightforwardly derived by taking the divergence of the
polynomials. The result is that the independent polynomial
coefficients are exactly specified by specifying the 
piecewise linear variation of the field components in the
transverse direction. A similar result obtains in three dimensions.
We are now in a position to specify the magnetic field at any point
in the interior of the square -- this being the problem of
divergence-free reconstruction. It is now easy to see that if the
square is subdivided into four squares through the refinement process
we can still specify the magnetic field at the faces of each of the
four smaller squares. Because the generating polynomial is
divergence-free, this specification is also divergence-free.
This solves the problem of divergence-free
prolongation from coarse meshes to fine meshes.
It is now easy to specify the four important algorithmic issues
in divergence-free AMR-MHD. They are:

{\bf (III.a) Time step sub-cycling on refined meshes:} This is 
entirely analogous to section II.a.

{\bf (III.b) Divergence-free prolongation of coarse mesh magnetic field to
fine mesh boundaries:} This is discussed in the two paragraphs above.
For three dimensions the problem becomes more complicated but it has
been worked out in detail by Balsara (2001c).

{\bf (III.c) Electric field correction at the fine-coarse interface:}
In section II.c we saw that whenever the fine and coarse meshes are
time-synchronized the fine and coarse mesh fluxes have to be made
consistent at the interface between the fine and coarse mesh. The
electric field plays a central role in the divergence-free evolution
of the magnetic field, just as the flux plays a central role in 
the evolution of higher order Godunov schemes. As a result, following
a style of reasoning that is entirely analogous to section II.c,
we have to make the electric field consistent at the interface.
This is illustrated in Fig 1 for the evolution of the z-component
of the magnetic field. Further details have been provided in
Balsara (2001c).

{\bf (III.d) Restriction of fine mesh solution to coarse mesh:}
In a direct analogy to
Section II.d, when the fine and coarse meshes are temporally
synchronized we replace the coarse
mesh magnetic fields by the area-averaged fine mesh magnetic fields.

\section{PARALLEL PROCESSING OF AMR HIERARCHIES}

The previous two sections have described the algorithmic
issues in divergence-free AMR-MHD in some detail. However, carrying out
an AMR calculation requires one to understand several further issues.
They can be illustrated by looking at Fig. 3 which shows an AMR-MHD
simulation of a supernova remnant. Figs. 3a and 3b show the log of the 
density and pressure variables. Figs. 3c and 3d show the Mach number and
the magnitude of the magnetic field. Fig. 3e shows the color coded levels
in the AMR hierarchy, blue being the base level grid, yellow being the
first level of refinement and red being the second level of refinement.
Fig. 3f shows the divergence of the magnetic field showing that
it remains within the bounds of machine accuracy. This allows us
to motivate several issues in supporting parallel AMR calculations,
all of which are catalogued in detail in Balsara and Norton (2001). The
issues are:

{\bf (IV.a) Object-oriented representation of the AMR hierarchy:} We see
from Fig. 3e that the meshes concentrate themselves in regions of strong
shock, which is what we desire from an AMR calculation. However, 
We see that each level of refinement is made of many small meshes,
all of which taken together form the AMR level. This requires a
high level of abstraction modeling so that each small mesh,
along with all the data structures that are needed for representing the
MHD solver on a mesh can be manipulated as a single unit. 
This is made possible by object-oriented methods which allow us to
encapsulate the data as well as the functions that change the data
as a single unit.

{\bf (IV.b) Load balancing of each level:} AMR computations are CPU
and memory intensive. As a result, it is important to carry them out
on parallel machines. However, it is difficult to know how to deal
out the different meshes to the different processors of a parallel
machine. This process of dealing the meshes out should be done in such a
way that each processor should has the same amount of computational
load. This is done via a load balancer. Several different load balancer
strategies are catalogued and inter-compared in Balsara and Norton (2001).

{\bf (IV.c) Parallel processing of the AMR hierarchy:} Parallel 
processing of the AMR hierarchy is tantamount to processing each
level in the AMR hierarchy in a load balanced fashion. 
Balsara and Norton (2001) have shown that this can be done using modern
parallel processing techniques along with the load balancer described above.

{\bf (IV.d) Solution-adaptive evolution of the AMR mesh hierarchy:} We see
that as the SNR shock moves through the mesh, the grids that make up the
levels have to change in response to the shock. This requires that one
is able to flag regions that need refinement and put down new refined
meshes where they are called for while removing old refined meshes
from regions that do not require refinement any longer. 
Such strategies were developed by Berger and Rigoutsis (1991) and
Balsara and Norton (2001) have shown that they are easily parallelized.

\section{TESTS and APPLICATIONS}

\begin{figure*}[t]
\centerline{\epsfysize=13cm\epsfbox{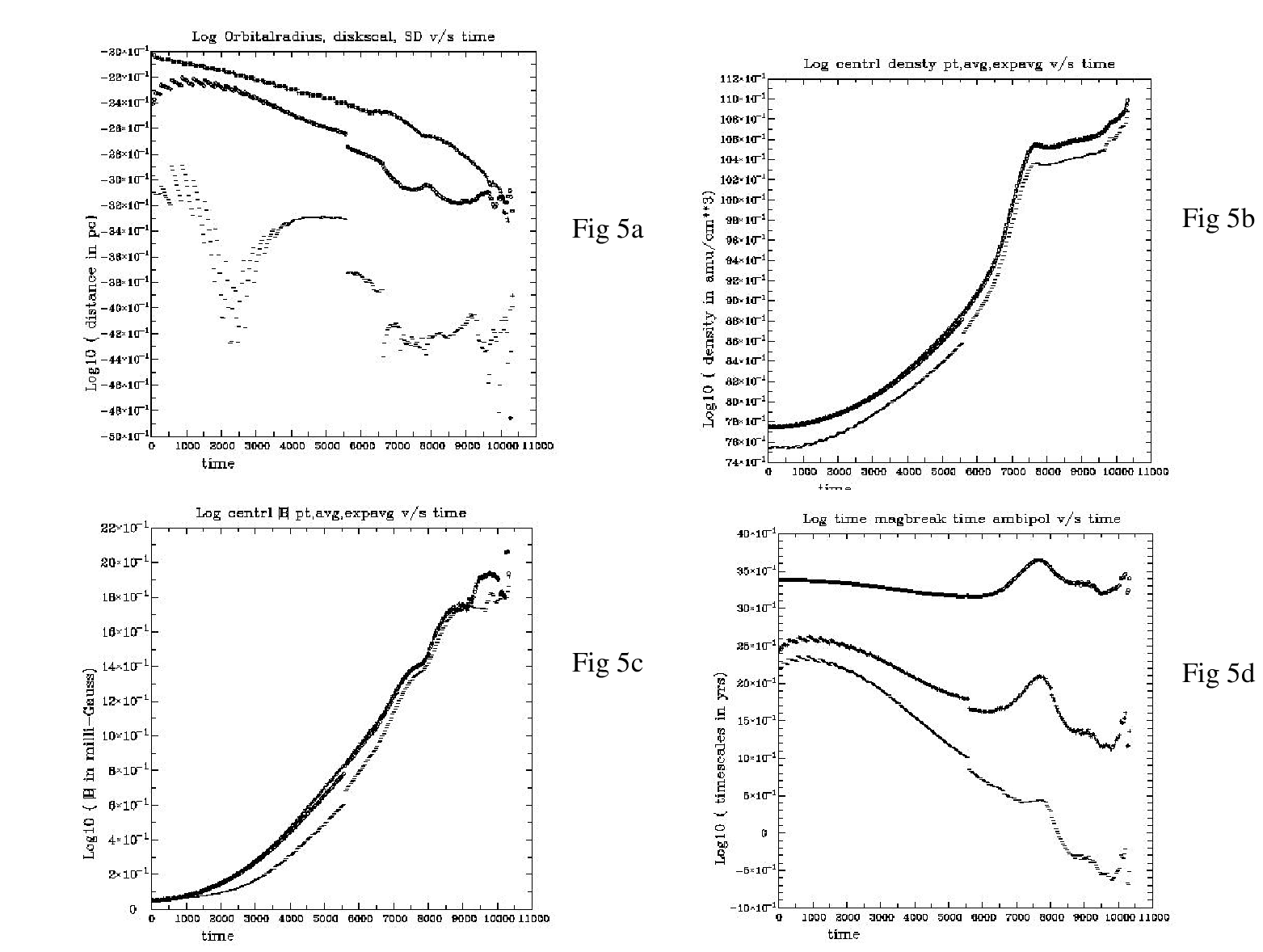}}
\vskip -0.5cm
\end{figure*}

\begin{figure*}[t]
\vskip -1.0cm
\centerline{\epsfysize=15cm\epsfbox{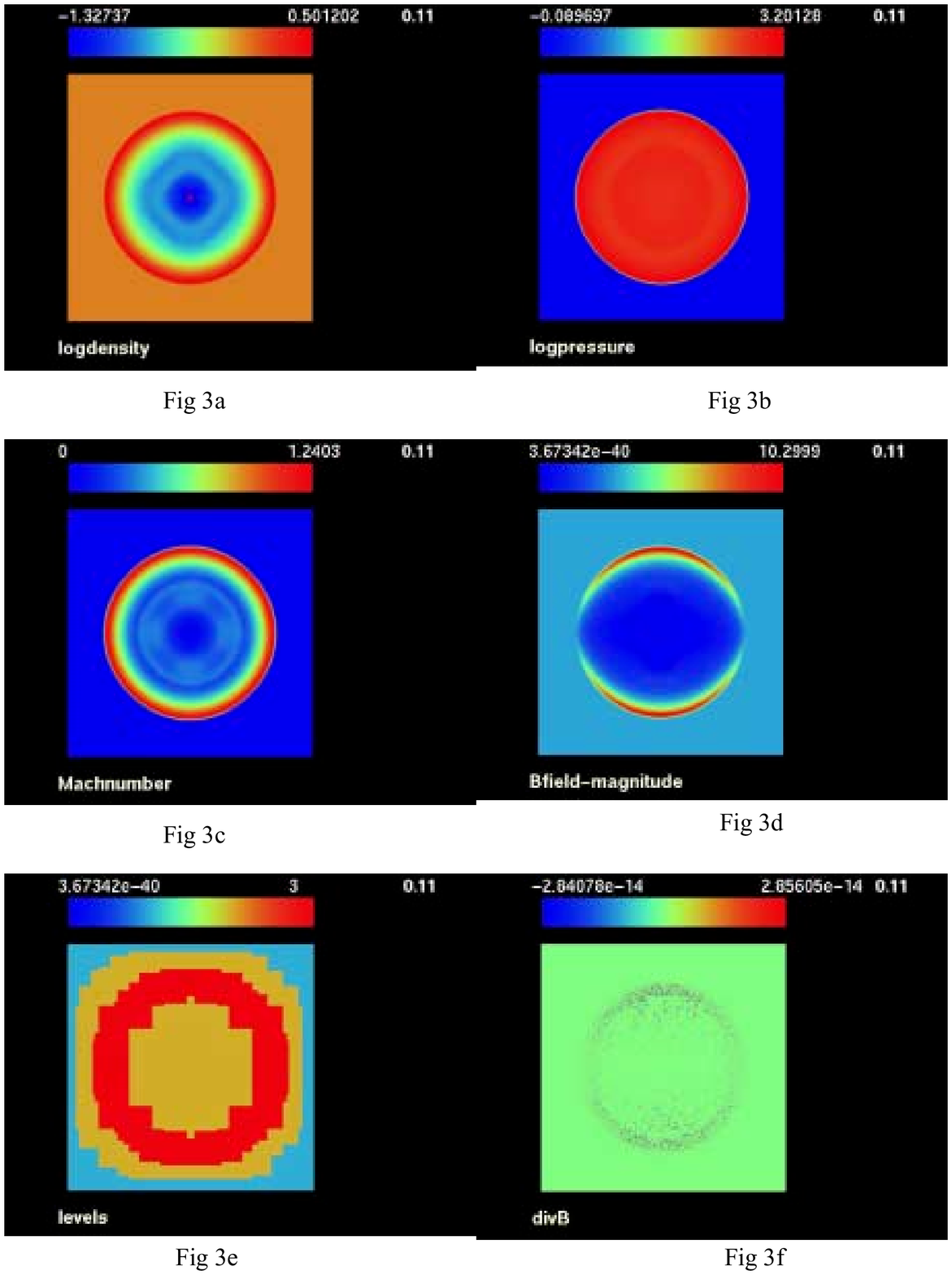}}
\vskip +17.5cm
\end{figure*}

{\bf (V.a) Test of Supernova Remnant Explosion:}
Balsara (2001c) provides several test problems to demonstrate
divergence-free AMR-MHD. The present test problem is motivated 
by the work of Balsara, Benjamin and Cox (2001) (hereafter BBC) 
who simulated the evolution of supernova remnants
and their propagation through the magnetized interstellar medium (ISM)
on large, i.e. $256^3$ zone, meshes. In this test problem we
do one of the same problems from BBC
using AMR-MHD on a $64^3$ zone base mesh and two levels of refinement.
Since the mesh refinement was carried out with a refinement ratio of
two across each level, the simulation with AMR has the same effective
resolution as the large uniform mesh simulation. Fig 3 shows the variables
from the AMR-MHD simulation. In Section IV we catalogued the variables
that were plotted out
in each of the sub-figures in Fig. 3. Similar variables from the large
uniform mesh simulation have been shown in BBC.
It can be seen that for the same effective resolution, the AMR-MHD
simulation produces the same quality of solution as the simulation
presented in BBC. This shows that there are some systems in computational
astrophysics where it is possible to intercompare the AMR-MHD simulation
with a large uniform mesh simulation and verify that the two simulations
produce the same quality of result if they have the same effective
resolution. This is a powerful demonstration of the saliency 
and effectiveness of AMR-MHD techniques in computational astrophysics. We also
notice from Fig. 3f that the divergence of the magnetic field has remained
within the bounds of machine accuracy!

\begin{figure*}[t]
\vskip -1.0cm
\centerline{\epsfysize=15cm\epsfbox{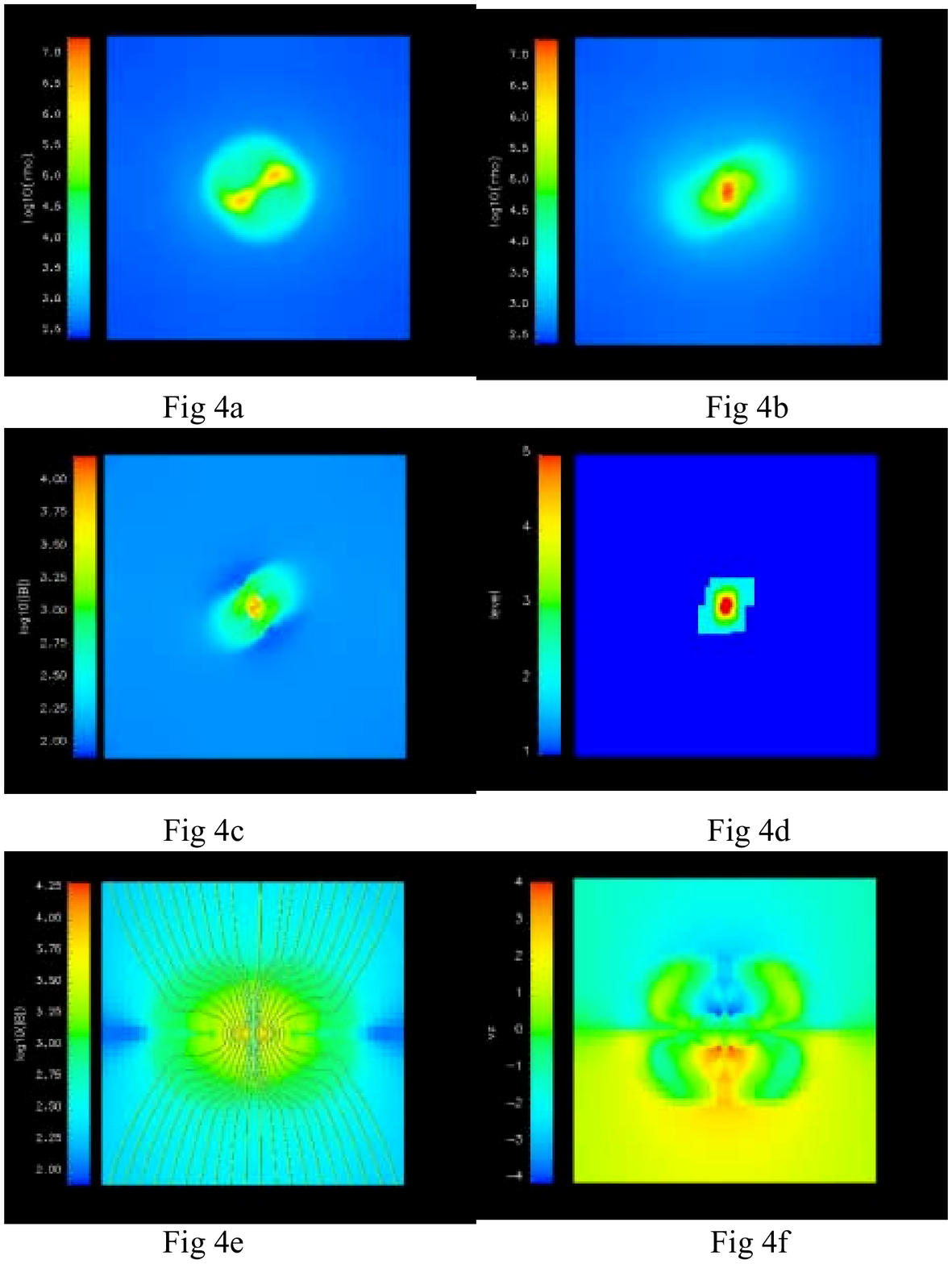}}
\vskip +17.5cm
\end{figure*}

{\bf (V.b) Application to Protostellar Core Collapse and Fragmentation:}
Understanding how protostellar cores form from a rotating density
condensation has long been a problem of great interest in star formation
studies. However, previous studies, see Truelove et al (1997) and
Boss et al (2000) have been hydrodynamical. The above authors have
shown that AMR is extremely useful for such simulations because 
the formation of density condensations is not correctly represented
when the mesh-based Jeans criterion is not met (which inevitably
happens in a uniform mesh calculation). I.e. if too much mass
is concentrated in one zone so that the zone becomes Jeans unstable then
the fluid flow is not properly represented on such a mesh. In such 
situations, our only recourse is to refine the mesh. The above-mentioned
hydrodynamical studies permit one to answer some questions. However, they
do not permit one to answer the important questions associated with the 
inclusion of magnetic fields: (1) How does the inclusion of magnetic
fields change the disks that form? (2) Does the magnetic field naturally
form an hourglass morphology? (3) How does the angular momentum evolve?
(4) Do outflows form? Can we understand their formation mechanism? (5) The
inclusion of fields introduces a competition between three timescales --
the free-fall timescale, the magnetic braking timescale and the ambipolar
diffusion timescale. How do those three timescales compete with each other?

Fig. 4 from Balsara and Burkert (2001) shows figures from an AMR-MHD
simulation of this problem on a $128^3$ zone base mesh with four further 
levels of refinement. The magnetic field was aligned with the rotation
axis which was taken to be the z-axis. The barotropic approximation of
Boss et al (2000) was used but ambipolar diffusion was not included.
Fig. 4a shows the log (density) in the xy-midplane at a relatively early time.
We see that the system fragments into two disks, just like in the 
hydrodynamical case. Fig. 4b shows the same variable at a late time. We now
see that the disks in the MHD simulation have lost angular momentum and
coalesced, unlike the hydrodynamical case. This can be explained via
mechanisms similar to the magnetic braking idea of Paleologou and
Mouschovias (1983). Fig. 4c shows the magnetic field in the xy-plane
at the same time as Fig. 4b demonstrating that the field has been dredged
in by the infall. The field, however, is connected to the ambient gas and
so imparts some of the two-disk system's angular momentum to 
the ambient gas. The ambient medium is
thus spun up at the expense of the two-disk system which eventually
inspirals and coalesces to form a single disk 
as shown in Fig. 4b. Fig. 4d shows a color coded representation of
the levels at the same time as Figs. 4b and 4c showing how the levels are
nested one within the other and how they track the increasing density.
Fig. 4e shows a zoomed view of the magnetic field lines in the
central part of the yz-midplane at a late
time. We see that the accretion processes have caused the field to
naturally organize itself into an hourglass morphology. The colors in
Fig. 4e track magnetic pressure. Fig. 4f shows the z-velocity in the
same portion of the yz-midplane. We see from the colors that an outflow
has been established. We claim that it is a magneto-centrifugal outflow
because it establishes itself in the same areas of the rotating gas
where the gas has a substantial rotational velocity and is restricted
to regions where the magnetic
field makes a large angle with the rotation axis. This is consistent
with the theory for MHD-outflows presented by Blandford and Payne (1982).

In Fig. 5 we show plots that make several aspects of the problem
more quantitative. Bacmann et al (2001) found that Class 0 cores
have a centrally flattened density profile. In keeping with their
observation, we fit the profiles
of either disk to a Gaussian profile $ \rho \sim exp [-(R/R_D)^2] $ where
$R$ is the radius of the disk and $R_D$ is the scale of the disk. 
The top curve in Fig. 5a shows the distance between the disks (in pc) as a
function of time (in years). The middle curve in Fig. 5a shows $R_D$
as a function of time. We see that at $\sim 7500 yrs$ the disks coalesce
resulting in a cusp in either of the two curves. The lower (and very jagged)
curve in Fig. 5a traces out the standard deviation in our measurement of
$R_D$ . We see that the standard deviation is substantially smaller than
our measurement of $R_D$ , indicating that our choice of a 
Gaussian density profile was a good one. ( We note, however, that 
the observations and simulations have latitude for other 
centrally flattened profile fits. However, any centrally flattened profile
would make allowance for a flattening length scale much like $R_D$ .)
It is worthwhile asking how the central density and magnetic field in
the disks evolve? The upper plots in Figs. 5b and 5c show the evolution
of the central density and magnetic pressure respectively 
in the disk/s as a function
of time. The lower plots in Figs. 5b and 5c measure the mean 
density and magnetic field in the disk/s averaged over the disk scale $R_D$ .
Again, the reorganization of material at the time of coalescence at
$\sim 7500 yrs$ is shown by a cusp in either plot.
While the central density can reorganize itself the central field cannot.
As a result, the field shows a steeper rise than the 
density after $\sim 7500 yrs$ .
It is always worthwhile asking how the inclusion of ambipolar drift (AD)
would change this scenario. Should AD operate, this
process of dredging in field would not be as pronounced. While we have
not included AD, we have evaluated its effect on
a post-facto basis. This is important because the 
the free-fall, the magnetic braking and AD
timescales compete in regulating the early evolution of
protostellar cores. This is shown in Fig. 5d where the upper plot
shows the magnetic braking time, the middle plot shows the ambipolar 
diffusion time for cosmic ray heating and the lower plot shows the
ambipolar diffusion time for far ultraviolet heating. The infall time
is a little larger than the magnetic braking time. We see that the
braking does dominate in this problem so that the disks coalesce
before infall to the center, as would be expected in such a situation.
However, with the inclusion of AD, it is possible for the AD timescales
to be much shorter suggesting that AD could reorganize the field structures
in times that are shorter than the infall time. To summarize, the
inclusion of magnetic fields has indeed shown the rich interplay of
new physics and timescales in this very important problem.

\section{CONCLUSIONS}

Several advances in divergence-free AMR-MHD and its application to
astrophysics are reported here. We
list them below:


(1) A general strategy is presented for the time-update of
the MHD system of equations on AMR hierarchies.

(2) Just as Berger and Colella (1989) reduced the conservative time-update of
the Euler equations on an AMR hierarchy to the application of a few simple
steps, we have reduced the divergence-free time-update of the MHD equations
on an AMR hierarchy to the application of a few simple steps. The steps
have been summarized in Section III.

(3) A significant advance has been made in the divergence-free
reconstruction of vector fields. 

(4) Divergence-free prolongation of magnetic fields on an AMR hierarchy can
be carried out via a very slight extension of the divergence-free
reconstruction scheme mentioned in the previous point.

(5) A divergence-free restriction strategy is presented.

(6) An electric field correction strategy is presented which restores the
consistency of electric fields at a fine-coarse interface in the AMR hierarchy.

(7) Because of the above four points, the time-step can be sub-cycled on
finer meshes without loss of the divergence-free property of the magnetic
fields.

(8) The above-mentioned innovations have been incorporated in the RIEMANN
framework for parallel, self-adaptive computational astrophysics. Several
stringent test problems have been presented and it is shown that the method
presented in this paper for AMR-MHD is truly divergence-free.

(9) Several very useful insights into astrophysical processes have
been derived from the AMR-MHD simulations that have been presented.


\acknowledgments{
This work was supported by NSF grants AST-0132246 and 005569-001.}


\begin{references}

\reference{} Bacmann, A. et al, A \& A, to appear (2001)

\reference{} Balsara, D.S., J. Comput. Phys., vol. 114, 284, (1994)

\reference{} Balsara, D.S., Ap.J.Supp., vol. 116, pg. 119, (1998a)

\reference{} Balsara, D.S., Ap.J.Supp., vol. 116, pg. 133, (1998b)

\reference{} Balsara, D.S., JQSRT, vol 61(5), 617, (1999a)

\reference{} Balsara, D.S., JQSRT, vol 61(5), 629, (1999b)

\reference{} Balsara, D.S., JQSRT, vol 62, 167, (1999c)

\reference{} Balsara, D.S., JQSRT, vol 61 (5),
637, (1999d)

\reference{} Balsara,D.S., JQSRT, vol 62, 167,
(1999e)

\reference{} Balsara, D.S. and Spicer, D.S., J. Comput. Phys., vol. 149, pg. 270, (1999)

\reference{} Balsara, D.S., vol. 132, pg. 1, Astrophys.J.Supp., (2001a)

\reference{} Balsara,D.S., JQSRT, vol 69(6), 671, (2001b)

\reference{} Balsara, D.S., to appear, J. Comput. Phys., (2001c)

\reference{} Balsara, D.S., and Norton, C.D., Parallel Computing, vol. 27,
pgs. 37-70, (2001)

\reference{} Balsara, D.S., Benjamin, R. and Cox, D., "The Evolution of Adiabatic
Supernova Remnants in a Turbulent Magnetized Medium", 2001, to appear,
Astrophys. J., vol. 563

\reference{} Balsara, D.S. and Burkert, A., "Three Dimensional AMR-MHD 
Simulations of Protostellar Core Collapse and 
Fragmentation", $198^{th}$ AAS Conf.

\reference{} Berger,M., and Colella,P., J. Comput. Phys., vol. 82, pp. 64, (1989)

\reference{} Berger,M., and Rigoutsos, I., IEEE Transactions on System, 
Man and Cybernetics, vol. 21, pp. 61-75, (1991)

\reference{} Blandford, R.D. and Payne, D.G., MNRAS, 199, 883, (1984)

\reference{} Boss, A.P., et al, Ap.J., 528, 325, (2000)

\reference{} Brackbill, J.U., and Barnes, D.C., J. Comput. Phys., vol. 35,
pg. 462 (1980)

\reference{} Brackbill, J., Space Sci. Rev.,
vol. 42, pg. 153 (1985)

\reference{} Brio, M. and Wu, C.C.,
J. Comput. Phys., v75, pp. 400-422, (1988)

\reference{} Dai, W., and Woodward, P.R., J. Comput. Phys., vol. 111, pg. 354, (1994)

\reference{} Dai, W., and Woodward, P.R., 
Astrophys. J., vol. 494, pg. 317, (1998)

\reference{} Londrillo, P., and Del Zanna, L., 
Astrophys. J., vol. 530, pg. 508,
(2000)

\reference{} Paleologou, M. and Mouschovias, T.C., Ap.J., 275, 838, (1983)

\reference{} Powell, K.G., 
ICASE Report No. 94-24, Langley VA, (1994)

\reference{} Powell, K.G., Roe,P.L., Linde, T.J., Gombosi, T.I., and DeZeeuw, D.L.,
 J. Comput.
Phys., vol. 154, pg. 284, (1999)

\reference{} Roe,P.L., and Balsara, D.S., 
SIAM J. Num. Anal., 56, 57, (1996)

\reference{} Ryu,D., and Jones,T., Ap. J., 442, 228, (1995)

\reference{} Ryu, D., Miniati, F., Jones, T.W., and Frank, A.,
Astrophys. J., vol. 509, pg. 244, (1998)

\reference{} Truelove, J.K. et al, Ap.J., 489, 179, (1997)

\reference{} VanLeer, B., J. Comput. Phys., v32, pp.101-136, (1979)

\reference{} Woodward, P., and Colella, P., J. Comput. Phys., v54, pp. 115-173, (1984)

\reference{} Zachary, A.L., Malagoli, A., and Colella, P., 
SIAM J. Sci. Comput., vol. 15, pg.
263, (1994)



\end{references}
\end{document}